\documentstyle[twoside,fleqn,espcrc2,graphicx]{article}

\newcommand{\AmS}{{\protect\the\textfont2
  A\kern-.1667em\lower.5ex\hbox{M}\kern-.125emS}}

\title{ SUSY/SUGRA/String phenomenology }

\author{Pran Nath\address{Department of Physics, 
        Northeastern University, \\ 
        Boston, MA 02115, U.S.A.}%
        \thanks{This research is supported in part by NSF grant
        number PHY-9602074.}
          }
       
\begin{document}

\begin{abstract}
A  review of  the current status of SUSY/SUGRA/String phenomenology 
is given.  

\end{abstract}

\maketitle

\section{INTRODUCTION}
The unification of gauge coupling constants observed in the 
extrapolation of the high precision LEP data points to the
general validity of the ideas of  supersymmetry/supergravity and 
grand unification.
This apparant success  also then 
validates  
the underlying  field theoretic approach at least up to the scale
$M_{GUT}\approx 10^{16}$ GeV. It is then entirely reasonable that one may 
extrapolate the grand unified theory  to the post 
GUT region perhaps all the way up  to the string scale 
$M_{string}$\cite{kaplu} where 
\begin{equation}
M_{str}=\frac{e^{(1-\gamma)/2}3^{-\frac{3}{4}}}{4 \pi}
 g_{string}  M_{Planck}\approx 5\times 10^{17}
\end{equation}
Thus $M_{string}$ appears to provide a natural cutoff 
 above which one is in the
domain of quantum gravity while below this scale physics may be described
by an effective field theory based on the principle of gauge invariance
and unified symmetry. 
 Consistency  of the two approaches is then ensured
by matching the grand unified theory  with the string at the string scale.
In this sense  the string serves to provide the
boundary conditions for the field theoretic approach at the 
scale $M_{str}$.

 Our objective of course is to eventually find the correct string model which 
 will probe physics in a continuous fashion from the very low energy 
 scales all the way up to the Planck scale. 
	However, among the many  problems to be overcome in 
	 string model building 
	is the enormity of the number of  possible vacua. For
	example, the ten dimensional $E_8\times E_8$ 
	heterotic string\cite{gross} 
	after compactification can generate models with
	rank up to 22. This yields many possibilities 
	and a variety of phenomenological models such as those based on free  
	field constructions,
	orbifolds, Calabi-Yau
	 compatifications etc have been investigated\cite{ibanez}.
\section{String Models}
		One way to limit the number of possibilities in
		string model building is to use the constraint that a GUT
		model exist below the string scale.
		As mentioned in the Introduction the LEP data appears to give 
		support to this picture.  Thus 
		it appears very appealing to build on the success of  
		SUSY/SUGRA  GUTS implied by the LEP data and try to  deduce
		a GUT model from strings. 
		Now the appearance of a gauge symmetry in strings 
		is characterized by world sheet currents whose
		operator product expansion defines the Kac-Moody level 
		k of the algebra:
		
		\begin{equation}
		j_a(z)j_b(w)\sim \frac{i f_{abc}}{z-w}j_c(w)+\frac{k}{2}
		\frac{\delta_{ab}}{(z-w)^2}+..
		\end{equation}		
		where  k is a positive integer for
		non-abelian groups while it is unconstrained  for the
		abelian case. 
		
		The earliest attempts at string model building were all
		within the framework of level 1. Here while one can
		achieve a unified symmetry with a desirable gauge 
		group\cite{lewellen}
		such as SU(5), SO(10) or E(6), one has the problem of
		not having massless adjoint scalars  
		which allow one to break the unified symmetry.
		While massless ajoint scalars can be gotten at  
		level 2, there are presently
		no known models which have 3 massless generations
		at this evel. For level
		3 one can generate models with 3 massless generations 
		as well as massless adjoint scalars.
		
		Over the last couple of years the construction of 
		level 3 models has been energetically pursued and models
		of the type SU(5), SO(10), and E(6) have been 
		constructed\cite{kaku1}.
		These models have many desirable features such as 
		3 chiral families, N=1 spacetime supersymmetry, massless
		 scalars in the adjoint representation and a non-abelian 
		 hidden sector. The phenomenology of one of the E(6)
		 models and of the related SO(10) model has also been partially
		 examined\cite{kaku2}. 
		 With the assumption that a non-perturbative
		 mechanism stabilizes the dilaton VEV, the gaugino 
		 condensation scale here is  
		  around $10^{13}$ GeV leading to a 
		  weak SUSY scale in the TeV region. 
		However, the adjoint Higgs is flat modulus, and the 
		mass matrix for the Higgs doublets is  rank 6 requiring
		a    fine tuning   to get a 
		    pair of massless Higgs doublets. 
		    Similarly, the texture for the up quarks is rank
		    3 rather than rank 1 as is desirable for  
		     the mass  hierarchy. Here
		  again  one needs a fine tuning for rank  reduction.
	         A satisfactory resolution to these may perhaps arise in 
	         other models of this generic type 
	         \cite{kaku1} whose phenomenology is not
	         yet fully investigated.  

	Of course, it is not  necessary to have a GUT  for the
	unification of the  gauge couplings $g_i$ 
	in strings as 
	the SM gauge group can emerge directly at the 
	string scale. Here one has\cite{ginsparg}
	\begin{equation}
	g_i^2k_i=g_{string}^2=(\frac{8\pi G_N}{\alpha'})
	\end{equation}
	where $k_i$ are the Kac-Moody levels of  the subgroups,
	 $G_N$  is the Newtonian constant and $\alpha'$  is the
	Regge slope. 
	 One consequence of this  possibility  is 
	that models of  this type  will in general
	possess fractionally charged neutral states unless the SM
	gauge group arises from an unbroken SU(5) at the
	string scale, or unless $k>1$\cite{schell}. 
	In models with fractionally
	charged states one must find a mechanism to either make them
	massive or confine them to produce bound states which carry 
	integral charges.

	We discuss now briefly the coupling constant unification in string 
	models and LEP data.  Using renormalization group one has in general 

	\begin{equation}
	\frac{16\pi^2}{g_i^2(M_Z)}=k_i \frac{16\pi^2}{g_{string}^2}
	+b_iln(\frac{M_{str}^2}{M_Z^2})+\Delta_i
	\end{equation}
	where $\Delta_i$ contain both stringy and non-stringy effects. 
	Firstly, one knows that if one uses the
	MSSM spectrum and runs the RG equations from the string scale down to 
	$M_Z$, then values of $sin^2\theta_W$ will differ from  
	experiment by many standard deviations. This situation can be
	corrected by either finding large threshold corrections, 
	or using non-standard Kac-Moody levels or finding 
	extra matter in vector like representations in the region between
	the string scale and the electro-weak scale. The last possibility 
	where one has extra matter at an intermediate scale currently appears 
	to be  the most promising one\cite{dienes,othertalks}.

	 Finally, one may speculate as to what the future possibilities for
	 string model building may be. Over the past two years we have seen
	 what is being called the second string revolution. This 
	 development concerns the fact that  one finds
	 that the five string theories (Type 1, Type 2A, Type 2B, 
	 SO(32) heterotic and $E_8\times E_8$ heterotic) are connected by 
	 dualities and it is conjectured that they all  arise from
	 a single eleven dimensional unified theory, the M theory, whose low 
	 energy limit is eleven dimensional supergravity. Thus, for example, 
	 it appears  that
	 the strongly coupled SO(32) heterotic string is the weakly  
	 coupled 10D typeI  string\cite{polchinski}, and  
	 that the dual
	 of  10D $E_8\times E_8$ heterotic string is the eleven dimensional
	  theory compactified on $S^1/Z_2$\cite{horava}. 
	 
	 We should see a new wave of model building which exploits the
	 power of dualities. Already some work has appeared at the 
	 level of model building where 
	 the strong coupling limit of the $E_8\times E_8$  heterotic 
	 string is exploited\cite{antoniadis,dudas}. 
	 Another application is in black hole thermodynamics 
	 and the partial success in the deduction of the Bekenstein-Hawking 
	 entropy/area 
	 law\cite{beken} from a microscopic viewpoint\cite{strom}. 
	 The analysis here involves the new string (D-brane)
	 degrees of freedom. Finally a remarkable new result concerns
	 the appearance of additional massless modes that can arise even in the
	 weak coupling limit of M theory resulting from an expansion 
	 of the   gauge group at specific points in the moduli 
	 space\cite{witten2}.
	 This phenomenon may lead to yet new possibilities for model 
	 building.   
\section{Supergravity Unification}
		It is reasonable  to conclude that while the string holds
	great many possibilities the correct string model still alludes 
	us. For this reason one must continue with a bottom up approach
	as well, and 
	an effective field theory is the correct framework for such an
	approach. The basic elements of this approach have been in place since
	the early eightees\cite{chams,applied,cremmer,bagger}.
	 We recall briefly what this approach constitutes.
	One assumes that below the string scale one has an effective 
	N=1 supergravity theory coupled to matter and gauge. The theory is
	defined in terms of three arbitrary functions. These  are the
	gauge kinetic energy function $f_{\alpha\beta}(Q_a,Q^a)$, the
	superpotential
	 $W(Q_a)$, and the Kahler potential K($Q_a,Q^a$) where $Q_a$
	are the matter fields and  $Q^a$ are 
	 their hermitian conjugates. The effective potential
	of this theory is given in terms of G which depends on a 
	combination of W and K: $G$=$\kappa^2$ $K$+
	$ln[\kappa^6|W|^2]$ where $\kappa\equiv1/M_{Planck}$.  
		
	In the effective N=1 supergravity the simplest approach for the
	unification  of the electro-weak and the 
	strong interactions and gravity
	is supergravity grand unification. In such a model supergravity can
	break in the hidden sector and the breaking communicated to the
	visible sector by gravity\cite{chams}. 
	Using the non-positive definite nature
	of the potential one can adjust the vacuum energy to be zero here.
	 The model generates soft SUSY breaking in the visible
	sector and the $\mu$ term can also be generated in the Kahler potential
	and  transferred to the superpotential via a Kahler transformation.
	 Further, soft SUSY breaking induces breaking of the electro-weak
	 symmetry\cite{chams} with the preferred mechnism being 
	 radiative breaking\cite{inou}
	 which explains one of the long standing puzzles regarding how
	 the electro-weak 
	 symmetry breaks in the Standard Model. In the  minimal version 
	 of the model the soft SUSY breaking is specified by just 5  
	 parameters\cite{chams,barbi,hall,nac0}
	 which reduce to  four parameters and one sign after the  
	 radiative ymmetry breaking of the electro-weak symmetry breaking
	 occurs. One may choose these to be
	 
	 \begin{equation}
	 m_0, m_{1/2}, A_0, tan\beta, sign(\mu).
	 \end{equation}
	 
	 A great deal of SUGRA phenomenology has been carried out over
	 the past few years based on this parameter space\cite{ross}(For a 
	 review  see ref.\cite{swieca}).
	 An interesting aspect of supergravity grand unification 
	 is that with R pariy invariance the lightest neutralino is
	 the LSP over  most of the parameter space of supergravity
	 unified models and thus a candidate for cold dark 
	 matter (CDM)\cite{lsp,arnowitt}.

	One may ask what leads to the soft SUSY breaking scale of 
	 $m_s\sim 10^2$ GeV. It has been suggested recently that such
	a scale could arise from an anomaly free R symmetry. The idea  here
	is that the requirement of anomaly cancellation requires 3 or 4
	hidden sector fields  and the U(1) quantum numbers of these 
	fields are large. Minimization gives VEV's to these fields which are
	O(1/10) of the Planck mass. Thus this small number with a large
	power can reduce the Planck mass down to the scale of $O(10^2)$
	GeV\cite{cd}.

	Our discussion above has focussed  on the gravity mediated breaking
	of supersymmetry. Another possibility which has been discussed 
	anew recently is the gauge mediated breaking of supersymmetry.
	Here one introduces 
	 	messenger fields in vector like representations  which  
	 	are chosen to transform non-trivially under the Standard
	 	Model  gauge group, and couple to the
	 	fields that break supersymmetry. The breaking of supersymmetry
	 	in the visible sector occurs at the one loop level 
	 	 for the gauginos and at the two loop level  for
	 	the  scalars\cite{dine}. However, 
	 	 	one of the problems that emerges 
	 	 	is that the Peccei-Quinn symmetry cannot
	 	 	be broken by gauge interactions, and one is forced
	 	 	to introduce non-gauge interactions to accomplish
	 	 	this breaking. 	
	 	 	Further, in models of this type it is the gravitino 
	 	 	rather than the lightest neutralino  which 
	 	 	is the LSP.
	         	Now the gravitino mass must lie below 
	         	$\sim 1$ KeV 
	         	in order that the gravitinos do not overclose the
	         universe\cite{pagels}. With  a mass in the
	         above range  the gravitino 
	         cannot be  a candidate for cold dark 
	         	matter (CDM). Thus unfortunately one of the 
	         	most attractive features of supersymmetry, that 
	         	it produces a candidate for cold dark matter for 
	         	free, is lost in gauge mediated breaking of 
	         	supersymmetry.

	Phenomenologically there are some important differences between
	the spectra of gravity mediated and gauge mediated breaking of
	supersymmetry\cite{dine,strumia}. 
	For a class of gauge mediated models one can write\cite{dine}
	
	\begin{equation}
	M_i(M_m)=\frac{\alpha_i(M_m)}{4\pi} M_0
	\end{equation}

	\begin{equation}
	m_a^2(M_m)=\eta C_{ia} M_i^2(M_m)
	\end{equation}
	where a is the matter field index, $C_{ia}$ are the Casimir co-efficients 
	for the field a, $M_m$ is the messenger scale, 
	and $\eta$ characterizes the various  models.
	$\eta<1$ for the minimal model where there is just one SUSY 
	breaking singlet, but one can generate $\eta >1$ for
	more general situations.
\section{SUSY/SUGRA GUT and LEP Data}
	 While the LEP data extrapolated to high scales with just the MSSM
	 spectrum exhibits unification to a good accuracy\cite{lang} 
	 a closer look
	 reveals  that the theoretical results lie about 2 std
	 higher for $\alpha_s$ than the LEP data\cite{bagger1,das}.
	  The 2 std effect could 
	 arise from a variety of  sources.  One possibility is 
	 Planck scale corrections\cite{hill,sarid,das}, since 
	 because of the proximity of the Planck
	 scale to the GUT scale there could be corrections 
	 of $O(M/M_{Pl})$. Such effects could manifest via corrections to
	 the gauge kinetic energy function.
	 Thus  inclusion of the
	 Planck  corrections brings in a field dependence in
	 $f_{\alpha\beta}$ and for 
	 SU(5) one may write 

	 \begin{equation}
	 f_{\alpha\beta} =(A\delta_{\alpha\beta}+
	  \frac{c}{2M_P} d_{\alpha\beta\gamma}\Sigma^{\gamma}) 
	 \end{equation}
         Here  c parameterizes Planck physics and  
         $c\sim 1$ gives the desired 2 std correction in the 
         gauge coupling constant unification to achieve full 
         agreement with experiment\cite{das}. 
 	 One can also understand a 2 std effect from extensions of the minimal
	 SU(5) model such as, for example, in some versions of the
	  missing  doublet model\cite{tamvakis}.

	  Next we discuss the issue of baryon instability and show that
	  this is a general problem in SUSY and string models alike.
	  First, all unified models have p instabiltiy via lepto-quark 
	  exchange. In addition SUSY theories may have rapid p decay via
	  lepton and baryon number violating dimension 4 operators. 
	  This type of p decay can be suppressed by the
	  imposition of R parity invariance. However, p decay can occur
	  via dimension 5 operators which is mediated by the Higgs triplet 
	  exchange\cite{wein,acn}. 
	  Thus, for example, in a model with n Higgs triplets 
	  and anti-triplets  one has in general an interaction such as 
	 
	 \begin{equation}
	 \bar H_1J + \bar K H_1 + \bar H_i M_{ij}H_j 
	 \end{equation}
	 
The condition for the suppression of p decay from dimension 5 operators then is 
that\cite{predictions}
      	   \begin{equation}
	 (M^{-1})_{11}=0
	 \end{equation}
	 \noindent
	 Such a suppression can arise either from discrete symmetries 
	 or from non standard embeddings. However, aside from these 
	 possibilities one finds that normally in SUSY/string models
	 one will in general have p instability and one must suppress 
	 dim 5 p decay by making the Higgs triplet superheavy, which
	 of course leads to the question of doublet-triplet splitting.
	 One of the interesting ideas in doublet-triplet splitting is
	 the use of a sliding singlet. It was shown some time ago 
	 \cite{sen} that the sliding singlet idea can be made to 
	 work in SU(6). 
	 However, in the analysis of ref.\cite{sen} one has the 
	 breaking of SU(6) to
	 SU(3)x SU(2) x U(1) occuring  in two steps which  
	 involves an intermediate scale of $\sim 10^{10}$ GeV and  
	 leads to a value of sin$^2\theta_W$  smaller than the
	 current experimental value. A recent analysis\cite{barr}
	  accomplishes the
	 breaking in one step eliminating the need for an intermediate scale.
\begin{figure}[htb]
\vspace{9pt}
\includegraphics [angle=270, width=2.9in]{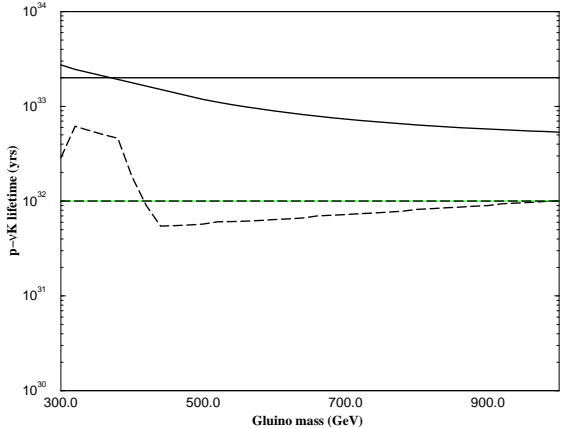}\\
\vspace{-1.0cm}
\caption{The maximum $p\rightarrow \bar\nu K^+$ lifetime in minimal supergravity
model under the constraint $m_0\leq 1$ TeV. The solid curve is for the case
when no relic density constraint is imposed.The dashed curve is with
the constraint $\Omega h^2<1$. The dashed horizontal line is the current
experimental limit and the solid horizontal line is the limit expected at  
Super-K and Icarus}
\label{fig:cutrelic}
\end{figure}
	    Recent theoretical analyses of p decay 	   
	 show that the expected Superkamiokande (Super-K)\cite{totsuka} 
	 and Icarus\cite{icarus} limit of $2\times 10^{32}$ yrs 
	 for the $p\rightarrow \bar \nu K^+$ mode can almost
	exhaust the maximum life time limits  for this mode in the  
         minimal SU(5) SUGRA models within the naturalness constraints of
        $m_0\leq 1$ TeV, $m_{\tilde g}\leq 1 $ TeV\cite{predictions}. 
        Including the constraints
        of dark matter one finds that the gluino mass must lie below 
        500 GeV if Super-K reaches its expected sensitivity and no
        p decay mode is seen\cite{oak}.\\
        
         The situation regarding SO(10) is more
        severe. In SO(10) models one has a large tan$\beta$, i.e.,
        tan$\beta\sim 50$ to get $b-t-\tau$ unification. However, the 
        effective  proton decay scale $M_{PD}\equiv (M^{-1})_{11}$ 
        to stabilize the proton
        is given by 
        \begin{equation}
        M_{PD}> tan\beta (0.57\times 10^{16})  GeV 
        \end{equation}
         which is $\sim 2.5 \times 10^{17}$ GeV. It is shown by Urano and
         Arnowitt that such a large scale 
	 upsets the $\alpha_s$ prediction of the minimal SUSY 
	model\cite{urano} and one needs large threshold corrections as in the
	work of Lucas and Raby  to restore 
	agreement with experiment\cite{lucas}.  
\section{ GUTS, Strings and Textures}
	We discuss now very briefly textures and their effects on proton
	 stability
	As is well known in unified theories one normally gets poor 
	predictions for the quark-lepton mass ratios. Thus, for example,
	in SU(5) $m_b/m_{\tau}$ is in good agreement with 
	experiment\cite{bbo,das} but
	$m_s/m_{\mu}$ and $m_d/m_e$ are not  in agreement and to get the
	correct mass ratios one needs 
	textures\cite{georgi,anderson,jain,binetruy,nathprl}.  
	In the Higgs doublet sector one may write 
	\begin{equation}
	W_{d}=H_1 l A^Ee^c+H_1d^c A^Dq+ H_2u^c A^Uq
	\end{equation}
	where $A^E$, $A^D$ and $A^U$ are the textures matrices.
	The simplest example of 
	textures are those by Georgi and Jarlskog where 
	$A^E$ and $A^D$ are given by

\begin{equation}
\left(\matrix{0&F&0\cr
                  F&-3E&0\cr
                  0&0&D\cr}\right),
\left(\matrix{0&Fe^{i\phi}&0\cr
                  Fe^{-i\phi}&E&0\cr
                  0&0&D\cr}\right)                 
\end{equation}	
respectively and  $A^U$ is given by	
\begin{equation}
\left(\matrix{0&C&0\cr
                  C&0&B\cr
                  0&B&A\cr}\right)
\end{equation}

	\noindent
	 where  the entries A,B,C,..
	 possess a hierarchy, i.e., A is O(1), B and D  are O($\epsilon$),
	 C and E are O($\epsilon^2$), and F is of order ($\epsilon^3$)
	 where $\epsilon <<1$. There are several interesting features 
	 regarding these textures. One of these concerns the origin of 
	 $\epsilon$ and several possibiliites have been discussed as to how
	 a small number such as this can naturally arise.
	 For example, it is suggested that the smallness of 
	 $\epsilon$ arises from the  ratio
	 $M_{str}/M_{Pl}$\cite{jain},  or in the
	 context of anomalous U(1) horizontal symmetries it is 
	 suggested that the ratio could be $<\theta>/M_{pl}$ where 
	 $\theta$ is a dynamical field which develops a VEV below the 
	 string scale\cite{binetruy}. 
	 There  is another possibilty and that is that
	 $\epsilon$ could be the ratio  $M_{GUT}/M_{pl}$. The last
	 possibility is the one we will focus on here. Such a possibility
	 arises naturally if one generates textures via Planck scale 
	 corrections, i.e., if one expands the potential in terms of higher
	 dimensional operators in powers of the field that develops  a
	 heavy VEV over the Planck mass\cite{anderson,nathprl}. 
	 After spontaneous breaking of 
	 the GUT symmetry one will generate entries in the quark-lepton 
	 lextures which are suppressed by powers of ($M_G/M_{Pl})$.
	 
		In the picture discussed above where the textures are
		generated via Planck scale corrections one can get the
		correct mass ratios with $\lambda_{yuk}\sim 1$. Of 
		course, once one fixes the textures in the Higgs doublet
		sector, the textures in the Higgs triplet sector can be
		computed. Now it turns out that if one does the most 
		general analysis with the Planck scale interactions,
		one can generate a large number of different sets of 
		textures in the Higgs triplet sector 
		for a given set of textures in the Higgs doublet
		sector\cite{nathprl}. Thus one needs a dynamical priniciple to 
		constrain the Planck scale corrections. One suggestion
		is to extend supergravity unification to include an 
		exotic sector\cite{nathprl}. 
		The fields in the exotic sector couple
		to fields in the hidden sector and to the would be 
		heavy fields of the visible sector. After spontaneous
		symmetry breaking the exotic fields become superheavy 
		and can be integrated out generating the Planck scale
		corrections. However, the Planck scale corrections
		are now  more constrained. If the fields in the 
		exotic sector are chosen to  be in the minimal vector 
		like representations, then the textures  in the 
		Higgs triplet sector are unique.
		Defining the textures in the Higgs triplet sector by
	\begin{eqnarray}	
	W_t=H_1 l B^E q + H_2 u^c B^U e^c\nonumber\\ 
	+\epsilon_{abc} (H_1 d^c_b B^D u^c_c + H_2^a u^c_b C^U d_c)
	\end{eqnarray}	
one gets\cite{nathprl}
		
\begin{equation}
B^E=\left(\matrix{0&aF &0\cr
  a^*F &{16\over 3}E &0\cr
                  0&0&{2\over 3}D\cr}\right) 
\end{equation}

\begin{equation}
B^D=\left(\matrix{0&-{8\over 27}F &0\cr
       {-8\over 27}F &-{4\over 3}E &0\cr
                  0&0&-{2\over 3}D\cr}\right)
\end{equation}

\begin{equation}
B^U=\left(\matrix{0&{4\over 9}C &0\cr
                  {4\over 9}C &0&-{2\over 3}B\cr
                  0&-{2\over 3}B&A\cr}\right)
\end{equation}
where a=$(-{19\over 27}+e^{i\phi})$  and $C^U=B^U$. An  estimate 
		shows that  inclusion of textures gives a moderate
		modification of the  decay branching ratios.  Further,
		the textures affect in a differential way the 
		various decay modes which in turn can be used to  
		provide a window on the textures at the GUT scale.
\section{Non-universality of Soft SUSY Breaking}
	Much of the analysis in supergravity unification is done within
	the framework of minimal supergravity unification with 4  parameters
	and one sign as discussed earlier. However, the framework of
	supergravity unification allows for non-universalities in the
	soft SUSY breaking sector of the theory. For example, a general 
	kinetic energy function will lead to a modification of the scaling
	laws for the chargino, the neutralino and the gluino masses. 
	In a similar
	fashion non-universalities can appear in the scalar sector of  the
	theory via a non-flat  Kahler potential\cite{soni,kapl,planck}. 
	 It is a reasonable
	proposition to relax the constraints on the minimality assumption
	to see  what effect one might have on low energy physics. Of course,
	there are severe constraints on non-universalities from flavor
	changing neutral currents (FCNC) which must be respected.  
	
	We begin with a discussion of non-universality in the gaugino
	sector. For the case of the general gauge kinetic energy function
	discussed earlier one has 
\begin{equation}
(m_{\frac{1}{2}})_{\alpha\beta}=\frac{1}{4} \kappa^{-1}\langle
G^a (K^{-1})^a_b~f_{\alpha\beta, b}^{\dag}\rangle m_{3/2}
\end{equation}	

\noindent
where  $G=\kappa^2 K+\ell n [\kappa^6\mid W \mid^2]$,
$G^a\equiv\partial G/\partial Q_a$ and($K^{-1})^a_b$ is the matrix
inverse of the Kahler metric $K_b^a$.
 For the case when $f_{\alpha\beta}$$\sim$
$\delta_{\alpha\beta}$ one has universal gaugino masses at the GUT scale,
i.e.,
\begin{equation}
M_i =(\alpha_i(Q)/\alpha_G)m_{1/2};~~~i = 1,2,3
\end{equation}
(Radiative corrections to this formula are dicussed in refs.\cite{mv}).
However, for the general $f_{\alpha\beta}$ case there would be important 
corrections. Here the corrections to the gaugino masses at the GUT scale 
involve not only the derivatives of the 
function $f_{\alpha\beta}$ but also the the
 derivatives of the Kahler potential.  Thus the corrections to the 
 gaugino masses are not identical to  the corrections to the
 gauge coupling constants arising from $f_{\alpha\beta}$. In general
 one has 
 
 \begin{equation}
 M_i=\frac{\alpha_i(Q)}{\alpha_G}(1+c'\frac{M}{M_{pl}}n_i) m_{\frac{1}{2}}
 \end{equation}
where $\alpha_i$ are the subgroup gauge coupling constants,
$\alpha_G$ is the GUT coupling constant,  
 $c'$ (which depends
on c) parametrizes the Planck scale correction, and 
$n_i=(2,-3,-1)$
 characterize the subgroups in the product $SU(3)\times SU(2)\times U(1)$.

For the universal case SUSY particles obey the scaling relations over 
most of the SUGRA parameter space under the constraint of electro-weak 
symmetry breaking. The scaling phenomenon arises because 
under the constraint of 
electro-weak symmetry breaking one finds that over most of the parameter 
space  one has  $\mu^2/M_Z^2>> 1$ which leads to the scaling  laws\cite{lsp}

\begin{eqnarray}
2m_{\chi_{1}}^0&\cong& m_{\chi_{1}}^{\pm}\cong m_{\chi_{2}}^0\simeq {1\over 3}
m_{\tilde g}\nonumber\\
m_{\chi_{3}}^0 &\cong& m_{\chi_{4}}^0 \cong m_{\chi_{2}}^{\pm}\simeq \mu >>
m_{\chi_{1}}^0\nonumber\\
m_{A}&\cong& m_{H^0} \cong m_H^{\pm} 
\end{eqnarray}
Inclusion of non-universlaities, however, will produce corrections to
the scaling laws.

	Next we discuss the scalar sector of the theory. Here the 
	non-universalities enter via the Kahler potential 
	$K(Q_a,Q^a)$ which in 
	general has the expansion 
	\begin{equation}
	K=\kappa^2K_0+K^a_b Q_aQ^b+(K^{ab}Q_a Q_b+ h.c.)+..
	\end{equation}
	Non-universalities appear when $K^a_{b,h}\neq 0$,
	$K^{ab},_h\neq 0$ where h is a  hidden sector field. As discussed
	in the next section current data on flavor changing neutral current
	(FCNC) processes impose impressive constraints on model building.  
	One of the sectors in 
	which the FCNC constraints are not that
	severe is the Higgs sector, and infact non-universalities in the
	Higgs sector are needed in SO(10) models with $b-t-\tau$ 
	unification to achieve radiative breaking of the electro-weak 
	symmetry\cite{matallio}.
	 However, it has been shown that if you allow for 
	non-universalities in the Higgs sector then you must allow for
	non-universalities also in the third generation sector as they are
	highly coupled because of the large top 
	Yukawa coupling\cite{nonuni,nwa}.
	It is convenient to parametrize the non-universalities in the 
	Higgs sector and in the third generation sector by \\
	
	\begin{equation}
	m_{H1}(0)^2=m_0^2(1+\delta_1),m_{H2}(0)^2=m_0^2(1+\delta_2)
	\end{equation}

	\begin{equation}
	m_{\tilde t_L}(0)^2=m_0^2(1+\delta_3),m_{\tilde t_R}(0)^2=
	m_0^2(1+\delta_4)
	\end{equation}
	where  a reasonable range  of $\delta_i$ is given by 
	$|\delta_i|\leq 1$.
	
	To exhibit the strong coupling of the Higgs sector and the third 
	generation sector we exhibit the corrections to the parameter 
	$\mu^2$ due to non-universalities\cite{nonuni}\\
	
	\begin{eqnarray}
	\Delta \mu^2=m_0^2(t^2-1)^{-1}(\delta_1-\delta_2 t^2
	-\frac{D_0-1}{2}\delta t^2)\nonumber\\+ {1\over 22} 
	{t^2+1\over t^2-1} S_0\biggl(
	1-{\alpha_1(Q)\over\alpha_G}\biggr)
	\end{eqnarray} 
\noindent
where 
$D_0\cong 1-m_t^2/(200 sin\beta)^2$ and 
\begin{eqnarray} 
S_0=Tr(Ym^2)=(m_{H_{2}}^2-m_{H_{1}}^2)\nonumber\\
+ \sum^{ng}_{i=1} (m_{\tilde q_L}^2 - 2
m_{\tilde u_R}^2+m_{\tilde d_R}^2-m_{\tilde l_L}^2 
+m_{\tilde e_R}^2)
\end{eqnarray}	
and $\delta\equiv (\delta_2+\delta_3+\delta_4)$. 
	From the above we see that $\delta_3$ and $\delta_4$ appear on
	an equal footing with $\delta_1$ and  $\delta_2$ indicating that
	non-universalities in the third generation sector must be
	included along with those in the Higgs sector for an appropriate
	treatment of these sectors. Non-universalities can affect 
	low energy phenomena such as analyses of dark matter and $R_b$.
	We shall discuss the effect of non-universalities
	on $R_b$ in Sec. 8.
\section{Constraints of  FCNC}
	 Suppression of FCNC  processes impose serious
	 constraints on model building.
	 First one has the well known constraint of the suppression of 
	 $K_s\rightarrow \mu^+\mu^-$ where the branching ratio is 
	 $<3.2 \times 10^{-7}$. In the SM this suppression is gotten via
	 the GIM mechanism where the  loop diagram allowing for this 
	 process is suppressed  by
	 $(m_c^2-m_u^2)	/M_W^2\approx 10^{-4}$.
	 In supersymmetric models there are additional diagrams where 
	 one has exchange of the chargino and squarks. Here the loop 
	 diagram  
	 contains a factor $(m_{\tilde c}^2-m_{\tilde u}^2)/M_{\tilde W}^2$.
	 Now in the minimal supregravity model with universal boundary 
	 conditions one has 
	  $(m_c^2-m_u^2)$ $\approx$ $(m_{\tilde c}^2-m_{\tilde u}^2)$. 
	   Thus the minimal supergravity unification
	  automatically generates the super GIM mechanism which again 
	  leads to a natural suppression of the  
	  $K_s\rightarrow \mu^+\mu^-$. Of course any inclusion of 
	  non-universalities in the SUGRA boundary conditions must
	  respect this constraint and that is what was done in
	  the analysis of non-universalities above. 
	  
	  There are of course also other FCNC processes which put
	  constraints on models. One of the more important of 
	  these is the  process $b\rightarrow s+\gamma$.
	 Recenlty the CLEO Collaboration\cite{alam}
	 has determined this branching ratio  to be 
	 \begin{equation}
	 BR(b\rightarrow s+\gamma)_{exp}=(2.32\pm0.67)\times 10^{-4}
	 \end{equation}
	 This process occurs at the loop level in the SM via the exchange
	 of the W and Z bosons and contains a significant QCD enhancement 
	 factor.
	 Including the  leading order and most of the next to leading 
	 order QCD corrections one finds that in the SM the 
	 $b\rightarrow s+\gamma$ branching ratio is given by 
	 \begin{equation}
	 BR(b\rightarrow s+\gamma)_{SM}=(3.48\pm 0.31)\times 10^{-4}
	 \end{equation}
	 for $m_t=176$ GeV.  
	 In supersymmetric models there are additional diagrams arising 
	 from the
	 exchange of  the charged Higgs, the charginos, the neutralinos 
	 and the gluino which contribute to this process\cite{bertolini}.
	 While the contribution from the charged Higgs exchange is always 
	 positive\cite{hewett} the sum of the remaining SUSY particle exchange 
	 contributions can be either positive or  
	 negative\cite{diaz}. In the minimal 
	 SUGRA 
	 model over most of the parameter space the charged 
	 Higgs turns out to be heavy, i.e., it has a mass much
	 larger than $M_Z$, and thus its contributions to 
	 $b\rightarrow s+\gamma$ 
	  is generally small. Among the
	 remaining SUSY particle exchanges it is the chargino exchange
	 which is  normally the largest contribution. Thus a large 
	 deviation of the experimental value of BR($b\rightarrow s+\gamma$)
	 from the SM value will point to the presence of a light chargino
	 mass  and a light stop mass. The constraint on the chargino and
	  the light stop mass depends of course on the  level
	  of deviation from the SM value. However, as a guideline one
	  can expect  that  the light chargino mass  and the 
	  light stop mass should
	  be in the vicinity of 100 GeV or below to make 
	 any significant contribution. The $b\rightarrow s+\gamma$
	 experiment puts a strong constraint on dark matter 
	 analyses\cite{dark} and on other SUSY phenomenology.  \\
	 
	 Finally we mention the FCNC process $\mu\rightarrow e+\gamma$
	 which can  arise in a 
	 variety of different ways in SUSY/SUGRA/string models\cite{muegamma}. 
	 In most such schemes
	 the process involves Yukawa couplings in the lepton sector and 
	 thus carries information on physics at the GUT scale and/or on 
	 physics at the 
	 string/Planck scales where one believes such 
	 interactions originate. 
\section{$R_b$ Status}
	In the SM $R_b$ has the value\cite{lepgroup} 
	\begin{equation}
	R_b^{SM}=0.2159,~~m_t=175 GeV
	\end{equation}
	The experimental value of $R_b$ has been been shifting 
	over the last 2-3 years. In 1995 values of $R_b$ as large as 3.5 std 
	above the SM value was  reported. Since then the value of $R_b$
	has moved down and in 1996 the LEP group reported  
	\begin{equation}
	R_b^{exp}=0.2178\pm 0.0011
	\end{equation}
	The result of Eq.(30) is only 1.8 $\sigma$ above the SM result. 
	Further, a
	more recent evaluation by the ALEPH group indicates no deviation
	from the SM value while the other three LEP groups  still 
	report an $R_b$ anomaly. It is instructive to  review briefly
	the status of $R_b$ in SUSY models. In MSSM there are 
	additional contributions to $R_b$ which involve the exchange
	of the charginos, the neutralinos, the gluino, and the 
	stops\cite{boulw,wells,garcia,chank}. Of these
	the chargino-stop exchange diagrams are the most important
	and here the dominant contributions arise when the light
	 stop is mostly right handed. In MSSM by fine tuning 
	 of  parameters one can get an $R_b$ correction as large as 
	 $0.0022-0.0028$. However, in minimal SUGRA one finds that this
	 correction is much smaller, i.e., it lies in the range 
	 $\Delta R_b^{max}=0.0002-0.0003$\cite{wang,dasg}. 
	 For  the case of supergravity
	 analyses with non-universalities one finds a maximal correction
	 of $0.0011$  for $\mu<0$ and a maximal correction of $0.0008$
	 for $\mu>0$. The $\Delta R_b$ corrections would  also
	 partially help bridge the gap between the low DIS value of 
	 $\alpha_s$ of ($0.116\pm 0.005$) and the high LEP  value of $\alpha_s$
	 of ($0.123\pm 0.006$). $\Delta R_b$ gives a correction
	 to the LEP value of $\alpha_s$ of 
	 $\Delta \alpha_s$=-4 $\Delta R_b$ which would result in a maximal
	 correction to  $\alpha_s$ of $\Delta \alpha_s$=-0.0044 for  
	 $\mu<0$ and of -0.0032 for $\mu>0$. Thus  the maximal 
	 $R_b$ correction can bridge the gap only half way between 
	 the DIS value and the LEP value\cite{shifman,erhler}.\\

	  For the non-universal SUGRA case  the maximal $\Delta R_b$ 
	  correction puts a stringent constraint on the sparticle 
	  spectrum. For instance, if one requires that the 
	   $\Delta R_b$ correction be greater than $\sim 0.0006$ which
	   at the current level of accuracy is $\sim \frac{1}{2}\sigma$ then 
	   the  supergravity constraints require that the light
	   chargino and the light stop should have masses below 100 GeV,
	   and the gluino mass must lie below 450 GeV(525 GeV) for $\mu<0$
	   ($\mu>0$).

\begin{figure}[htb]
\vspace{9pt}
\includegraphics [angle=90, width=2.9in]{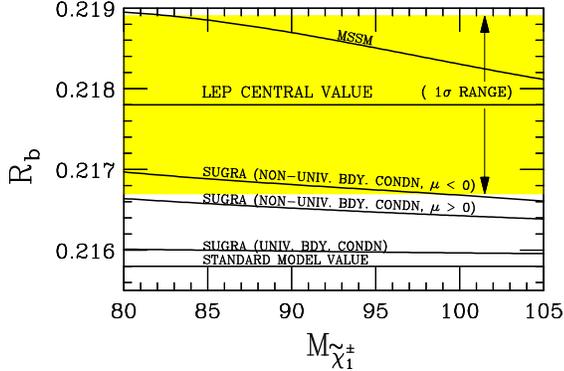}\\
\vspace{-1.5cm}
\caption{ $R_b^{max}$ as a function of the  light chargino mass for the 
Standard Model, for SUGRA model with universal boundary conditions and
for the SUGRA model with non-universal boundary conditions for  
 $\mu>0$ and $\mu<0$ from ref.[76].}
\label{fig:largenenough}
\end{figure}
{\small
\begin{tabular}{|l|c|c|}
\hline
 Quantity& Numerical Values \\ 
\hline 
$R_b^{exp}-R_b^{SM}$ & 0.0019$\pm$ 0.0011\\
$\Delta R_b^{SUSY(max)}$(MSSM) & 0.0022\cite{chank}\\
$\Delta R_b^{SUSY(max)}$(MSSM) & 0.0028\cite{dasg} \\
$\Delta R_b^{SUSY(max)}$(mSUGRA) & 0.0002\cite{dasg}\\
$\Delta R_b^{SUSY(max)}$(nSUGRA) & 0.0011($\mu<0$)\cite{dasg}\\
$\Delta R_b^{SUSY(max)}$(nSUGRA) & 0.0008($\mu>0$)\cite{dasg}\\
\hline
\end{tabular}
\\}

\noindent
 Table 1: The current experimental value of 
$\Delta R_b$ vs the maximal $\Delta R_b^{SUSY}$ in MSSM and
in SUGRA models with universal boundary conditions at the
GUT scale (mSUGRA) and with non-universal boundary conditions
(nSUGRA) from ref.\cite{dasg}.
\vspace{0.5cm}

 Further, in this situation the light Higgs must have a mass
 which lies below 93 GeV. Now at the Tevatron in the Main Injector
	   Era one will be able to probe the light chargino via the
	   trileptonic signal\cite{trilep} 
	    and the light
	   stops in the ranges indicated above. At TEV33 one 
	   will be able to explore gluino masses up to 450 GeV with an
	   integrated luminosity of  100$fb^{-1}$\cite{kamon,amidei}. 
	   Further, Higgs mass 
	   in the range indicated will be accessible at LEP II  if one
	   can achieve center of mass energies of $\sqrt s =192$ GeV.
	   A Higgs mass in the range above will also be accessible at
	   TEV33 with an integrated luminosity of 5-10$fb^{-1}$. This 
	   means that the non-universal SUGRA model with 
	   $\Delta R_b>0.0006$ can be completely tested in the 
	   chargino, stop, and Higgs sector at LEPII and at TEV33 and  
	   also fully(partially) tested in the gluino sector for 
	   the case $\mu<0(\mu>0)$ at at TEV33. 
	   Thus the results of non-universal
	   SUGRA analysis are very strong. One is predicting that
	   if there is any sizable $\Delta R_b$ correction which we
	   construe here to imply that $\Delta R_b>0.0006$, or
	$\Delta R_b>\frac{1}{2}\sigma$, then one must see either a
	chargino, or a stop and a Higgs at  LEPII and at TEV33. 
	The gluino must also be seen if $\mu<0$ while for $\mu>0$
	it could escape detection at TEV33 if it lies in the
	region between 450-525 GeV.
\section{The Brookhaven $g_{\mu}-2$ Experiment as a Probe 
	of SUGRA/String Unification}
	 $g_{\mu}-2$ is  a powerful probe of SUGRA/string unified theories.
	 We discuss briefly now as to the implications of the improved 
	 results on  $g_{\mu}-2$ expected from Brookhaven in the
	 near future\cite{kino}. 
	 The theoretical value of  $a_{\mu} \equiv (g_{\mu}/2		   	
 	-1)$ within the SM is given by 
 	
 \begin{equation}
  a_{\mu}^{theory}(SM)= 11659172(15.4) \times
 	10^{-10}
 \end{equation}	
 	 Here the theoretical value is computed to the $\alpha^5$
 	Q.E.D. order, and to  $\alpha^2$ order in the hadronic corrections,
 	and includes  two loop corrections from the electro-weak sector.
	The hadronic  correction is given by\cite{eidelman}
		  
\begin{equation}	
	 a_{\mu}^{hadron}(SM)= 687.0(15.4) \times 10^{-10}
\end{equation}	

\noindent 
while the electro-weak correction is given by\cite{czar}  

\begin{equation}	
 a_{\mu}^{EW}(SM)= 15.1(0.4) \times 10^{-10}.
\end{equation}	 

\noindent
	  Recently Alemany et.al.\cite{alemany} have used the new 
	 $\tau$ data at LEP  to achieve a significant 
 	reduction in the hadronic error, i.e., they get
 	
 \begin{equation}	
 	a_{\mu}^{hadron}(SM)= 701.4(9.4) \times 10^{-10}
 \end{equation}	
        One expects that the error in the hadronic
 	correction will  reduce further in the near future from
 	experiments at  VEPP-2M, DA$\Phi$NE, and 
 	BEPC\cite{kinoshita,dafne,bepc}. A reduction 
 	in the error by another factor of 2 is not out of reach.  
	The current experimental value of $a_{\mu} \equiv (g_{\mu}/2		   	
 	-1)$ is $a_{\mu}^{exp}= 11659230(84) \times 10^{-10}$ and
 	the new Brookhaven experiment is expected to reduce the 
 	uncertainty by a factor of 20\cite{hughes}. The expected
 	 reduction in the experimental error at Brookhaven and 
 	 a corresponding reduction in the hadronic error as discussed
 	 above will allow one to test the Standard Model electro-weak 
 	 contribution. However, it was pointed out early on that 
 	 any experiment that can test the SM electro-weak contribution
 	 will also test the supersymmetric contribution as often
 	 the supersymmetric contribution tends to be as large or larger
 	 than the SM contribution\cite{yuan}. The more 
 	 recent precision analyses of 
 	  $g_{\mu}-2$ within SUGRA\cite{lopez,chatto} and in 
 	  MSSM\cite{moroi} support these observations.
 \section{Test of GUT, Post GUT and String Physics via
	  Precision Measurement of Sfermion Masses}
 	  One of the interesting aspects of SUGRA unified models 
	  and string models is that sensitive measurement of 
	  sparticle masses can be used as a probe of physics
	  at post GUT and string scales. It is already known
	  that  sfermion masses carry information on gauge 
	  symmetry breaking at the GUT scale\cite{kawamura}. For example, the
	  breaking of SO(10) via various sequences leads to distinguishable
	  patterns in the sparticle spectrum. 
  	  It was recently suggested
	  that sparticle spectrum can also act as a probe of physics
	  in the post GUT region all the way up to the string 
	  scales\cite{snow,planck}.
	  For example, it is easy to demonstrate that  post GUT 
	  scenarios such as SU(5), SO(10), SU(3)$^3$, $G_{SM}$ etc.,
	  produce distinct signatures in the sfermion spectrum. 
	  
	  	We discuss now briefly how the various post GUT 
	  	scenarios can be tested. Consider for instance a 
	  	scenario where the group in the post GUT region is SU(5).
	  	In this case one finds that mass differences 
	  	$m_{\tilde e_L}^2-m_{\tilde d_R}^2$, 
	  	$m_{\tilde u_L}^2-m_{\tilde e_R}^2$, 
	  	$m_{\tilde u_L}^2-m_{\tilde u_R}^2$ are independent
	  	of the specific assumption of universality and
	  	thus  would vanish when extrapolated to the GUT scale.
	    One can also determine the string scale experimentally from
	    data on sfermion masses. In the above scenario and under
	    the assumption of universality of the soft SUSY parameters 
	    at the string scale, we can determine the string scale using
	    renormalization group  from the intersection, for example, of 
	    the lines  of $\tilde e_L$ and $\tilde u_L$ as a function of
	    running scale Q. Thus $M_{str}$ which is one of the most fundamental 
	    parameters of string theory can be determined from 
	    accurate data on the sfermion masses. A  similar analysis
	    can be carrried out for other post GUT possibilities. 
	    For example, consider the case of SO(10) in the post GUT region
	    with fermions in the three generations of 16-plets of SO(10)
	    where a 16 plet decomposes to 10+$\bar 5$+1 of SU(5), 
	    and the usual 5+$\bar 5$ of SU(5) Higgs lie in the 10 of 
	    SO(10).  Now assume that at $M_G$ one has breaking of SO(10) into 
	    the SM gauge group. In this case because of rank reduction
	    one has  D term contributions in the matching conditions
	    below $M_G$\cite{drees,kawamura}. 
	    Absorbing $\delta_{10}$ in the definition
	    of $\tilde m_0^2$ and  defining $m_5^2=\tilde m_0^2(1+\delta_5)$,
	     $m_{H_1}^2=\tilde m_0^2(1+\delta_1)$ 
	 	and $m_{H_2}^2=\tilde m_0^2(1+\delta_2)$ one has  
	 	\begin{equation}
	 	\delta_5=\delta_1-\delta_2
	 	\end{equation}
	    Thus SO(10) models of the type described above have an extra 
	    constraint which again can be tested by extrapolation of
	    the sfermion data from the low energy scales to the GUT
	    scale. Similar considerations can be extended to other  
	    post GUT scenarios such as SU(3)$^3$ and SU(3)$\times$SU(2)
	    $\times$U(1), and one finds that in each case accurate 
	    data on sfermion masses allows one to distinguish that 
	    particular scenario from others.
	
		One can extend the same general procedure to testing
		string models. The origin of supersymmetry breaking 
		in string theory is not fully understood and thus one 
		must rely on certain parametrizations\cite{font,munos}. 
		One useful approach in this direction is parametrizing soft 
		SUSY breaking by the VEVs of the dilaton (S), and the
		moduli fields ($T_i,U_i$). As an illustration we consider (2,2)
		Calabi-Yau compatifications  with the gauge groups 
		$E_6\times E_8$. Although the models investigated 
		thus far in this type of compactification have not
		resulted in any realistic model, the framework does
		provide a testing ground for some of the ideas discussed
		here. Thus we consider the case of a campactification
		with a single modulus T\cite{candelas} where the soft SUSY
		breaking parameters are universal at the string 
		scale\cite{munos}. In this case one may parametrize
		soft SUSY breaking by 
		
		\begin{eqnarray}
		m_{\frac{1}{2}}=\sqrt{3} sin\theta e^{-i\gamma_S} 
		m_{\frac{3}{2}}\nonumber\\
		A_0=-(1+\omega e^{-i(\gamma_T-\gamma_S)}cot^2\theta)
		m_{\frac{1}{2}}\nonumber\\
		m_0^2=\frac{1}{3}(1+\Delta e^{-i\gamma_S} cot^2\theta)
		m_{\frac{1}{2}}^2
		\end{eqnarray}
		where  $\theta$ is the angle between the dilaton and the
		Goldstino directions, $\omega$ and $\Delta$ include the
		sigma model and instanton contributions, and $\gamma_S$
		and $\gamma_T$ are CP violating phases. 
		On the CP preserving manifolds one has $\gamma_S=0=\gamma_T$
		and eliminating $\theta$ one gets a  constraint between
		the  SUSY breaking parameters 
	         $m_0$, $m_{\frac{1}{2}}$, and $A_0$, and  
	         $\Delta$ and $\omega$.
	         One has 
	         \begin{equation}
	         \omega^2/\Delta=(A_0/m_0+  m_{\frac{1}{2}}/m_0)^2
	         \end{equation}                   
	         Now the soft SUSY parameters $m_0$, $m_{\frac{1}{2}}$ 
	         and $A_0$ can be gotten from the sparticle mass 
	         measurements. These determinations then allow us to
	         test a given string model. Thus, it is shown in
	         ref.\cite{planck} that the SUSY parameters $M_2=120$ GeV,
	         $m_0=187$ GeV, $A_t=-285$ GeV, and tan$\beta=5$ rule
	         out the model with $\Delta=1.62$ and 
	         $|\omega |$=0.64\cite{munos}.
	         
	  The  discussion above shows that one can extract SUSY parameters 
	  from the sparticle
	  masses and use them to test specific string models. Of course 
	  the extraction of the SUSY parameters from the collider data
	  depends on the accuracy with which mass measurements  can be
	  made. Several recent papers have addressed this issue.
	  Hadron colliders 
	  may  be able to provide us with the mass measurements of
	  sparticle masses with accuracies of few percent\cite{baer,hinch}.
	  At linear colliders  one may be able to achieve
	  accuracies of up to 1-2$\%$ level\cite{tsukamoto,feng,kuhlman}. 
	  Such data can then be used to explore physics beyond the GUT scale.   
\section{Conclusion}
Supersymmetric models currently provide an attractive framework
for the solution to the hierarchy problem. 
Supergravity grand unification with spontaneous 
breaking of supersymmetry via  a hidden sector provides a concrete model 
where supersymmetric particle spectrum can be computed and their
interactions analysed. The minimal SUGRA model is
consistent with the LEP data with a possible 2 std discrepancy in 
$\alpha_s$, which may point to the existence of Planck scale  corrections.
Thus in addition to SUGRA unification we may already be witnessing 
Planck scale corrections in the LEP 
data. The minimal SUGRA also predicts that supersymmetric corrections 
to $R_b$ are small and that $R_b$ is close to the SM value.
 The  recent  experimental data on $R_b$ appears to be moving towards 
 eliminating the previously large ($\sim 3\sigma$) $R_b$ anomaly. This
 is precisely what is predicted by the minimal SUGRA model. 
 There are many other predictions of SUGRA models which are 
 testable in accelerator and non-accelerator experiments. These 
 include SUGRA corrections to  $b\rightarrow s+\gamma$, $g_{\mu}-2$,
 predictions on proton decay, and the existence of a low energy 
 supersymmetric particles spectrum which
 should be visible at colliders.
 
        An interesting interface of GUTs and strings would occur if
        SUGRA GUT can indeed  arise from strings. Recent progress
        in string GUTS with Kac-Moody levels $k>1$ appears encouraging
        from this view point and the models that arise here have
        many appealing features.
 However, the detailed phenomenology of all models of this type needs to
  be fully worked out to determine if a model satisfying all
 the phenomenological constraints will survive. 
 Another development which will shape phenomenology in the years ahead 
 is the development of string dualities and M theory. This development has  
opened new directions and perhaps set new ground rules for model
building. Perhaps enhanced symmetry groups and stringy effects at scales
far below the string scale may play a role in a new generation of model
building. However, prudence requires that in view of the enormity of
the problems at hand, i.e., the existence of many string vacua,
 difficulty regarding the breaking of supersymmetry in strings, 
 the problem of getting zero vacuum energy after SUSY breaks etc.,
 that it is desirable to  work from both ends, i.e.,
 from the top down as  in string theory and from  the bottom up as in
 SUSY/SUGRA  phenomenology. In either case 
  a hint from experiment regarding the existence of SUSY in nature 
   will be helpful. One hopes such a hint will come in the near future  from
   the various experiments underway, such as the Brookhaven 
   $g_{\mu}$-2 experiment, and CLEO's analysis of the process
 $b\rightarrow s+\gamma$, 
 from the  detection of an LSP in dark matter  detectors,  or in the direct  
 observation of supersymmetric particles at the Tevatron, at LEP or at 
 the LHC.


\begin{thebibliography}{9}
\bibitem{kaplu}
V. Kaplunovsky, Nucl. Phys. {\bf B307}(1988)145; E: ibid. {\bf B382}(1992)436.

\bibitem{gross} 
D. J. Gross, D.J. Harvey, E. Martinec, and R. Rohm, Phys. Rev. Lett. 
{\bf 54} (1985) 502; Nucl. Phys. {\bf B256}(1985) 253; ibid. {\bf B267}
(1986) 75. 
\bibitem{ibanez}
For a sample of string phenomenology see, 
L.E.Ibanez, H. P. Nilles and F. Quevedo, Nucl. Phys.{\bf B307}(1988)109;
A. Antoniadis, J.Ellis, J. Hagelin, and D.V. Nanopoulos, Phys. Lett. 
{\bf B194}(1987)231; B. R. Green, K. H. Kirklin, P.J. Miron G.G. Ross,
Nucl. Phys.{\bf B292} {1987}606; R. Arnowitt and P. Nath, Phys. Rev.
{\bf D40} (1989)191;  A. Farragi, Phys. Lett. {\bf B278} (1992)131;
M. Cvetic and P. Langacker, Phys. Rev. {\bf D54}(1996)3570.

\bibitem{lewellen}
D.C. Lewellen, Nucl. Phys. {\bf B337}(1990)61; J. A. Schwarz, Phys. Rev.
Phys. Rev. {\bf D42}(1990)1777; S. Chaudhuri,
S.-W. Chung, G. Hockney, and J.D. Lykken, Nucl. Phys.{\bf 452}(1995)89;
G.B. Cleaver,Nucl. Phys. {\bf B456}(1995)219.

\bibitem{kaku1}
Z. Kakushadze and S.H.H. Tye, Phys. Rev. {\bf D55}(1997)7896; ibid,
{\bf  D56}{1997)7878;  Phys. Lett. {\bf 392}(1997)325. 
\bibitem{kaku2}
Z. Kakushadze, G. Shiu,  S.H.H. Tye, and Y. Vtorov-Karevsky, 
hep-ph/970413.


\bibitem{ginsparg}
P. Ginsparg, Phys. Lett. {\bf B197}(1987)139.

\bibitem{schell}
A. Schellekens, Phys. Lett. {\bf 237}(1990)363.

\bibitem{dienes}

See, e.g., K. Dienes, "String Theory and the Path to Unification: 
A Review of Recent Developments", hep-ph/9602045.

\bibitem{othertalks}
For a discussion of other aspects of string phenomenology see talks by
I.Antoniadis, G. Cleaver, E. Dudas, J.-R. Espinosa,  Z. Kakushadze, 
F. Quevedo, and Y.-Y. Wu. 

\bibitem{polchinski}
J. Polchinski and E. Witten, Nucl. Phys.{\bf B460} (1996) 525.

\bibitem{horava}
P. Horava and E. Witten, Nucl. Phys. {\bf B460}, 506(1996).

\bibitem{antoniadis}
I. Antoniadis and M. Quiros, Phys. Lett. {\bf B392}(1997)61;
hep-ph/9705037.

\bibitem{dudas}
E. Dudas and  J. Mourad, hep-ph/9701048. 

\bibitem{beken}
J.D. Bekenstein, Phys. Rev. {\bf D12}(1975)3077; S.W. Hawking,
Phys. Rev. {\bf D13}(1976)191.

\bibitem{strom}
A. Strominger and C. Vafa, Phys. Lett. {\bf B379}(1996)99;
J. Maldacena and A. Strominger, Phys. Rev. Lett.{\bf 77}(1996)428;
C.V. Johnson, R.R. Khuri and R.C. Myers, Phys. Lett. {\bf B378}(1996)78.

\bibitem{witten2}
E. Witten, Nucl. Phys. {\bf B460}(1996)541.

\bibitem{chams}
A.H. Chamseddine, R. Arnowitt and P. Nath, Phys. Rev. Lett. {\bf 49}
(1982)970.
\bibitem{applied}
P. Nath, R. Arnowitt and A.H. Chamseddine, ``Applied N =1
Supergravity'' (World Scientific, Singapore, 1984).

\bibitem{cremmer}
E. Cremmer, S. Ferrara, L.Girardello and A. van Proeyen,
Phys. Lett. {\bf 116B}(1982)231.

\bibitem{bagger}
E. Witten and J. Bagger, Nucl. Phys. {bf B222}(1983)125. 

\bibitem{inou}
K. Inoue et al., Prog. Theor. Phys. {\bf 68}(12982)927; L. Iban$\tilde n$ez and
G.G. Ross, Phys. Lett. {\bf B110} (1982) 227; L. Alvarez-Gaum\'e, J. Polchinski
and M.B. Wise, Nucl. Phys. {\bf B221} (1983)495; J. Ellis, J. Hagelin, D.V.
Nanopoulos and K. Tamvakis, Phys. Lett. {\bf B125} (1983) 2275; L. E. Iba$\tilde
n$ez and C. Lopez, Phys. Lett. {\bf B128} (1983) 54; Nucl. Phys. {\bf B233}
(1984) 545; L.E. Iba$\tilde n$ez, C. Lopez and C. Mu$\tilde n$oz, 
Nucl. Phys. {\bf B256} (1985) 218.

\bibitem{barbi}
R. Barbieri, S. Ferrara and C.A. Savoy, Phys. Lett. {\bf B119} (1982) 343.
\bibitem{hall}
 L. Hall, J. Lykken and S. Weinberg, Phys. Rev. {\bf D27} (1983) 2359.
 \bibitem{nac0}
  P. Nath, R. Arnowitt and A.H. Chamseddine, Nucl. Phys. {\bf B227}
  (1983) 121.

\bibitem{ross}
G. Ross and R.G. Roberts, Nucl. Phys. {\bf B377} (1992) 571; R. Arnowitt and P.
Nath, Phys. Rev. Lett. {\bf 69} (1992) 725; M. Drees and M.M. Nojiri, Nucl.
Phys. {\bf B369} (1993) 54; S. Kelley {\it et. al.}, Nucl. Phys. {\bf B398},
(1993) 3; M. Olechowski and S. Pokorski, Nucl. Phys. {\bf B404} (1993) 590; G.
Kane, C. Kolda, L. Roszkowski and J. Wells, Phys. Rev. {\bf D49} (1994) 6173;
D.J. Casta$\tilde n$o, E. Piard and P. Ramond, Phys. Rev. {\bf D49} 
(1994) 4882; W. de Boer, R. Ehret and D. Kazakov, Karlsruhe preprint, IEKP-KA/94-05
(1994); V. Barger, M.S. Berger, and P. Ohmann, Phys. Rev. {\bf D49}
(1994) 4908; H. Baer, M. Drees, C. Kao, M. Nojiri and X. Tata, Phys.
 Rev. {\bf D50},
 (1994) 2148; H. Baer, C.-H. Chen, R. Munroe, F. Paige and X. Tata, Phys. Rev.
{\bf D51} (1995) 1046.  

\bibitem{swieca}
 R. Arnowitt and P. Nath, Proc. of VII J.A. Swieca Summer School
ed. E. Eboli (World Scientific, Singapore, 1994).

\bibitem{lsp}
R. Arnowitt and P. Nath, Phys. Rev. Lett. {\bf 69} (1992) 725;
P. Nath and R. Arnowitt, Phys. Lett. {\bf B289} (1992) 368.

\bibitem{arnowitt}
See also  talks by R. Arnowitt and J. Wells.

\bibitem{cd}
A. Chamseddine and H. Dreiner, hep-ph/9607261.



\bibitem{dine}
See  talks by M. Dine, C. Kolda, H. Murayama, S. Martin, A. Nelson,
N. Polonski, E. Poppitz, and S. Thomas and the references quoted there in.

\bibitem{pagels}
H. Pagels and J. Primack, Phys. Rev. Lett. {\bf 48}(1982)223.


\bibitem{strumia}
A. Strumia, hep-ph/9705306.



\bibitem{lang}
P. Langacker, Proc. PASCOS 90-Symposium, Eds. P. Nath and S. Reucroft (World
Scientific, Singapore 1990); J. Ellis, S. Kelley and D.V. Nanopoulos, Phys. Lett.
{\bf B249}(1990)441; {\bf B260}(1991)131; U. Amaldi, W. de Boer and H.
Furstenau, Phys. Lett. {\bf B260}(1991)447; F. Anselmo, L. Cifarelli, A.
Peterman and A. Zichichi, Nuov. Cim. {\bf 104A}(1991)1817; {\bf 115A}
(1992)581; P. Langacker and N. Polonski, Phys. Rev. {\bf D47}(1993)4028.
\bibitem{bagger1}
J. Bagger, K. Matchev and D. Pierce, Phys. Lett {\bf B348}(1995)443; 
P.H. Chankowski, Z. Plucienik, and S. Pokorski, Nucl. Phys. {\bf B439}
(1995)23.

\bibitem{das}
 T. Dasgupta, P. Mamales and P. Nath, Phys. Rev.
{\bf D52}(1995)5366; D. Ring, S. Urano and R. Arnowitt, Phys. Rev. {\bf D52},
(1995)6623; S. Urano, D. Ring and R. Arnowitt, Phys. Rev. Lett. {\bf 76}
(1996)3663; P. Nath, Phys. Rev. Lett. {\bf 76}(1996)2218.


\bibitem{hill}
C.T. Hill, Phys. Lett. {\bf B135}(1984)47; Q. Shafi and C. Wetterich, Phys.
Rev. Lett. {\bf 52}(1984)875.


\bibitem{sarid}
L.J. Hall and U.
Sarid, Phys. Rev. Lett. {\bf 70}(1993)2673; P. Langacker and N. Polonsky, Phys.
Rev. {\bf D47}(1993)4028.

\bibitem{tamvakis}
A. Dedes and K. Tamvakis, IOA-05-97; J. Hisano, T. Moroi, K. Tobe and
T. Yanagida, Phys. Lett. {\bf B342}(1995)138.

\bibitem{wein} S. Weinberg,~Phys. Rev. {\bf D26} (1982) 287; 
N. Sakai and T. Yanagida, Nucl. Phys.{\bf B197}(1982) 533; 
S. Dimopoulos, S. Raby  and F. Wilczek, Phys.Lett.
 {\bf 112B} (1982) 133;
J. Ellis, D.V. Nanopoulos and S. Rudaz, Nucl. Phys.
{\bf  B202} (1982) 43;
B.A. Campbell, J. Ellis and D.V. Nanopoulos,
 Phys. Lett. {\bf 141B} (1984) 299;
S. Chadha, G.D. Coughlan, M. Daniel
 and G.G. Ross, Phys. Lett.{\bf 149B} (1984) 47.

\bibitem{acn}
R. Arnowitt, A.H. Chamseddine and P. Nath, Phys. Lett.
{\bf 156B} (1985) 215;
P. Nath, R. Arnowitt and A.H. Chamseddine, Phys. Rev. {\bf 32D} (1985) 2348;
 J. Hisano, H. Murayama and T. Yanagida, Nucl. Phys.
{\bf B402}(1993) 46.

\bibitem{predictions}
R. Arnowitt and P. Nath, Phys. Rev. {\bf D49} (1994) 1479.


\bibitem{sen}
A. Sen,  Phys. Lett. {\bf B148}(1984)65; Phys. Rev. {\bf D31}(1985)900. 
\bibitem{barr}
S.M. Barr, hep-ph/9705266.

\bibitem{totsuka} Y.Totsuka, Proc. XXIV Conf. on High Energy Physics,
Munich, 1988,Eds. R.Kotthaus and J.H. Kuhn (Springer Verlag, Berlin, 
Heidelberg,1989).

\bibitem{icarus} 
Icarus Detector Group, Int. Symposium on Neutrino 
Astrophysics, Takayama. 1992.


\bibitem{oak}
P. Nath and R. Arnowitt, Proc. of the Workshop "Future Prospects of 
Baryon Instability Search in p-Decay and $n\bar n $ Oscillation 
Experiments", Oak Ridge, Tennesse, U.S.A., March 28-30, 1996, ed:
S.J. Ball and Y.A. Kamyshkov, ORNL-6910, p. 59.

\bibitem{urano}
S. Urano and R. Arnowitt, hep-ph/9611389.

\bibitem{lucas}
V. Lucas and S. Raby, Phys. Rev. {\bf D54} (1996) 2261; ibid. {bf D55}
(1997) 6986.


\bibitem{bbo}
V. Barger, M.S. Berger, and  P. Ohman, Phys. Lett. {\bf B314},
(1993) 351; W. Bardeen, M. Carena, S. Pokorski, and C.E.M. Wagner,
Phys. Lett. {\bf B320}(1994) 110.

\bibitem{georgi}
H. Georgi and C. Jarlskog, Phys. Lett. {\bf B86} (1979) 297;
J. Harvey, P. Ramon and D. Reiss, Phys. Lett. {\bf B92},
(1980) 309; P. Ramond, R.G. Roberts,
G.G. Ross, Nucl. Phys. {\bf B406}(1993) 19; L. Ibanez and
G. G. Ross, Phys. Lett. {\bf 332}(1994) 100; 
K.S. Babu and R. N. Mohapatra, Phys. Lett.{\bf 74}(1995) 2418;
K. S. Babu and S.M. Barr, hep-ph/9506261;
N. Arkani-Hamed, H.-C. Cheng and L.J. Hall, Phys. Rev. {\bf D53}
(1996} 413;  R.D. Peccei and K. Wang, Phys. Rev. {\bf D53}(1996)2712.

\bibitem{anderson}
 G. Anderson, S. Raby, S. Dimopoulos, L. Hall, and G.D.
Starkman, Phys. Rev. {\bf D49}(1994) 3660.

\bibitem{jain}
V. Jain and  R. Shrock, Phys. Lett. {\bf B35}(1995) 83.
\bibitem{binetruy}
P. Binetruy and P. Ramond, Phys. Lett. {\bf  B350}(1995) 49.

\bibitem{nathprl}
P. Nath, Phys. Rev. Lett. {\bf 76}(1996) 2218; 
Phys. Lett. {\bf B381}(1996) 147.


\bibitem{soni}
S. Soni and A. Weldon, Phys. Lett. {\bf B126} (1983) 215.

\bibitem{kapl}
V.S. Kaplunovsky and J. Louis, Phys. Lett. {\bf B306} (1993) 268.

\bibitem{mv}
S.P. Martin and M.T. Vaughn, Phys. Lett. {\bf B318} (1993) 331; D. Pierce and
A. Papadopoulos, Nucl. Phys. {\bf B430}(1994) 278.


\bibitem{matallio}
D. Matalliotakis and H. P. Nilles, Nucl. Phys. {\bf B435}(1995)115;
M. Olechowski and S. Pokorski, Phys. Lett. {\bf B344}(1995)201;
N. Polonski and A. Pomerol, Phys. Rev.{\bf D51}(1995)6532.


\bibitem{nonuni}
P. Nath and R. Arnowitt, hep-ph/9701301.
\bibitem{nwa}
P. Nath, J. Wu and R. Arnowitt, Phys. Rev. {\bf D52} (1995)4169.

\bibitem{alam}
M. S. Alam et. al. (CLEO Collaboration), Phys. Rev. Lett.{\bf 74},2885(1995). 
\bibitem{buras}
A. J. Buras, M. Misiak, M. Munz and S. Pokorski, Nucl. Phys. {\bf B424}
(1994)374;  M. Ciuchini et. al., Phys. Lett. {\bf B316}(1993)127.
\bibitem{chet}
K. Chetyrkin, M. Misiak, and M. Munz, Report. No. hep-ph/9612313. 

\bibitem{bertolini}
S. Bertolini, F. Borzumati and A. Masiero, Phys. Rev. Lett. {\bf 59},
(1987)180; R. Barbieri and G. Giudice, Phys. Lett. {\bf B309}(1993)86. 

\bibitem{hewett}
J.L. Hewett, Phys. Rev. Lett. {\bf 70}(1993)1045; V. Barger, M. Berger,
P. Ohmann. and R.J.N. Phillips, Phys. Rev. Lett. {\bf 70}(1993)1368.

\bibitem{diaz}
M. Diaz, Phys. Lett. {\bf B304}(1993)278; J. Lopez, D.V. Nanopoulos,
and G. Park, Phys. Rev. {\bf D48}(1993)974; R. Garisto and J.N. Ng,
Phys. Lett. {\bf B315}(1993)372; J. Wu, R. Arnowitt and P. Nath,
 Phys. Rev. {\bf D51}(1995)1371; 
 V. Barger, M. Berger, P. Ohman and R.J.N. Phillips, Phys. Rev. {\bf D51},
 2438(1995); H. Baer and M. Brhlick, hep-ph/9610224.


\bibitem{dark}
P. Nath and R. Arnowitt, Phys. Lett. {\bf B336}(1994)395;
  F. Borzumati, M. Drees, and M.M. Nojiri, Phys. Rev.{\bf D51},
 (1995)341.



\bibitem{muegamma}
For recent analyses see, R. Arnowitt and P. Nath, Phys. Rev. Lett. 
{\bf 66}(1991)2708; R. Barbieri, L. Hall and A. Strumia, Nucl. Phys.
{\bf B445}(1995)219; H. Goldberg and M. Gomez, Phys. Rev. {\bf D53}
(1996)5244.
 
 
\bibitem{lepgroup} The LEP Electroweak Working Group, LEPEWWG/96-01(1996).


 \bibitem{boulw} M. Boulware and D. Finnel, Phys.Rev. {\bf D44}
 (1991)2054;
A. Djouadi, et. al., Nucl.Phys.
{\bf B349}(1991)48;
A. Djouadi, M. Drees and H. Konig, Phys.Rev.{\bf D48}(1993)3081.

\bibitem{wells} J.D. Wells, C. Kolda and G. Kane, Phys.Lett.{\bf B338}
(1994)219; J. Wells and  G. Kane, Phys. Rev. Lett. {\bf 76}(1996)869.
 
\bibitem{garcia}
 D. Garcia, R. Jimenez, and Sola, Phys. Lett. {\bf B347}, 
(1995)321; E. Ma and D. Ng, Phys. Rev. {\bf D53}(1996)255;
A.Brignole, F.Feruglio, and F. Zwirner,Z.Phys. {\bf C71} (1996)679.
D. Garcia and J. Sola,Phys. Lett. {\bf B354} (1995)335;


\bibitem{chank}
 P.H. Chankowski and S. Pokorski, Nucl. Phys. {\bf B475},
(1996)3; M. Drees, R.M. Godbole, M. Guchait, S. Raychaudhuri,
and D.P. Roy, Phys. Rev. {\bf D54}(1996)5598.

\bibitem{wang}
 X. Wang, J. Lopez and D.V. Nanopoulos,Phys. Rev. {\bf D52}
 (1995)4116.

 \bibitem{dasg}
 T. Dasgupta and P. Nath, hep-ph/9702442 (to appear in Phys. Rev.D).
 
 
\bibitem{shifman}
M. Shifman, Mod. Phys. Lett. {\bf A10}(1995)605.
\bibitem{erhler}
 J. Erhler and P. Langacker, Phys. Rev. {\bf D52}(1995)441.


\bibitem{trilep}
P. Nath and R. Arnowitt, Mod. Phys. Lett. {\bf A2}(1987)331;
R. Barbieri, F. Caravaglio, M. Frigeni, and M. Mangano, Nucl. Phys.
{\bf B367}(1991)28; H. Baer and X. Tata, Phys. Rev. {\bf D47},
(1992)2739;  J.L. Lopez, D.V. Nanopoulos, X. Wang and A. Zichichi,
Phys. Rev. {\bf D48}(1993)2062; H. Baer, C. Kao and X. Tata, Phys.
Rev. {\bf D48}(1993)5175.
 

\bibitem{kamon}
T. Kamon, J. Lopez, P. McIntyre and J.J. White, Phys. Rev.{\bf D50}
(1994)5676; H. Baer, C-H. Chen, C. Kao and X. Tata, Phys. Rev. 
{\bf D52}(1995)1565; S. Mrenna, G.L. Kane, G.D. Kribbs, and T.D. Wells,
Phys. Rev. {\bf D53}(1996)1168. 


\bibitem{amidei}
D. Amidie and R. Brock, " Report of the tev-2000 Study Group",\\ 
FERMILAB-PUB-96/082.



\bibitem{kino}
T. Kinoshita and W. J. Marciano, in "{\it Quantum Electrodynamics}",
edited by T. Kinoshita (World Scientific, Singapore, 1990).pp. 419-478. 

\bibitem{eidelman}
S. Eidelman and F. Jegerlehner,
 Z. Phys. {\bf C67}(1995)585; See also D. H. Brown and
W.A. Worstell, Phys. Rev. {bf D54}(1996)3237.

\bibitem{czar}
A. Czarnicki, B. Krauss and W. Marciano, TTP95-19; hep-ph/9506256.

\bibitem{alemany}
R. Alemany, M. Davier and A. Hocker, hep-ph/9703220. 
\bibitem{kinoshita}
T. Kinoshita, talk at the 10th International Symposium on High Energy 
Spin Physics, Nagoya, Nov. 9-14, 1992.
\bibitem{dafne}
R. Barbieri and E. Remiddi, in the $DA\Phi NE$ Physics Handbook, ed.
L. Maiani, G. Pancheri, and N. Paver, INFN, Frascati(1992)301.
\bibitem{bepc}
Status report of BEPC and BES, Z. Zheng and N. Qi (CCAST World lab,
Beijing, and Beijing Inst. High Energy Phys.), 1989; Tsukuba 1989,
Proceedings $e^+e^-$ collision physics, 55-73. 
\bibitem{hughes}
See, e.g., V. Hughes, Proc. of PASCOS-91, eds. P. Nath and S. Reucroft
(World Scientific), p. 868; T. Hasegawa et. al. (Universal Academy Press,
Tokyo, 1992), p.717.
\bibitem{yuan}
T. C. Yuan, R. Arnowitt, A.H. Chamseddine and P. Nath, Z. Phys. {\bf C26},
(1984)407;
D. A. Kosower, L. M. Krauss, N. Sakai, Phys. Lett. {\bf 133B}(1983)305.
\bibitem{lopez}
J. Lopez, D.V. Nanopoulos, and X. Wang, Phys. Rev. {\bf D49}(1994)366.
\bibitem{chatto}
U. Chattopadhyay and P. Nath, Phys. Rev. {\bf D53}(1996)1648;
  M. Carena, G.F. Giudice
and C.E.M. Wagner, CERN-TH/96-271.

\bibitem{moroi}
T. Moroi, Phys. Rev. {\bf D53}(1996)6565.


\bibitem{kawamura}
Y. Kawamura, H. Murayama and M. Yamaguchi, Phys. Lett. {\bf B324}
(1994)52; H. Murayama, M. Olechowski and S. Pokorski, Phys. Lett.
{\bf B371}(1996)57.


\bibitem{snow}
R.~Arnowitt and P.~Nath, in 
"Physics and Technology of the NLC: Snowmass 96", 
ed. by S. Kuhlman et. al., 
hep-ex/9605011.

\bibitem{planck}
R.~Arnowitt and P.~Nath, hep-ph/9701325
(to appear in Phys. Rev. D).


\bibitem{drees}
M. Drees, Phys. Lett. {\bf B181}(1986)279; P. Nath and  R. Arnowitt, 
Phys. Rev. {\bf D39}(1989)2006; J.S. Hagelin and S. Kelley, Nucl. Phys.
{\bf B342}(1990)95.

\bibitem{font}
A. Font, L.E. Ibanez, D. Lust and F. Quevedo, Phys. Lett. {\bf B245} (1990)
401; M. Cvetic, A. Font, L.E. Ibanez, D. Lust and F. Quevedo, Nucl. Phys.
 {\bf B361} (1991) 194; A. de la Macorra and G. G. Ross, Nucl. Phys. 
 {\bf B404} (1993) 321; V. Kaplunovsky and J. Louis, Phys. Lett. {\bf B306}
 (1993)269; R. Barbieri, J. Louis and M. Moretti, Phys. Lett. {\bf B312}
 (1993)451; (Err. {\bf B316} (1993) 632); J. Lopez, D.V. Nanopoulos, and 
 A. Zichichi, Phys. Lett. {\bf B319} (1993)451; S. Ferrara, C. Kounnas 
 and S. Zwirner, Nucl. Phys. {bf B429} (1994) 589 (Err. {\bf B433}
 (1995) 255).
 
 \bibitem{munos}
 H. Kim and C. Munos, hep-ph/9608214.

\bibitem{candelas}
P. Candelas, M. Lynker and R. Schimmrigk, Nucl. Phys. {\bf B341}(1990)
383; J. Fuchs, A. Klemm, C. Scheich  and M.G. Schmidt,
Phys. Lett.{\bf B232}(1989)317.

\bibitem{baer}
H. Baer, C. Chen, F. Paige, and X. Tata, Phys. Rev. {\bf D52}(1995)2746.

\bibitem{hinch}
I. Hinchliffe, F.E. Paige, M.D. Shapiro, J. Soderqvist and
W. Yao, Phys. Rev.{\bf D55} (1997)5520.

\bibitem{tsukamoto}
T. Tsukamoto, K. Fujii, H. Murayama, M. Yamaguchi, and Y. Okada,
Phys. Rev. {\bf D. 51}(1995)3153.
\bibitem{feng}
J.L. Feng, M.E. Peskin, H. Murayama, and X. Tata, Phys. Rev. {\bf D52}
(1992)1418.

\bibitem{kuhlman}
S. Kuhlman et. al., 
"Physics and Technology of the NLC: Snowmass 96", hep-ex/9605011.
\end{thebibliography}
\end{document}